\documentclass[12pt]{article}

\title{Symmetries, Symmetry Breaking, Gauge Symmetries\thanks{Talk at the Triennial International Conference ``New Developments in Logic and Philosophy of Science'', Rome 18-20 June, 2014}}

\author{F. Strocchi \\ INFN, Sezione di Pisa, Pisa, Italy}
\date{}

\makeglossary

\usepackage{latexsym}
\usepackage[T1]{fontenc}
\usepackage{amsmath}
\usepackage{amssymb}

\def \AO {{\cal A}({\cal O})}
\def \AO' {{\cal A}({\cal O}')}

\def \be {\begin{equation}}
\def \ee {\end{equation}}
\def \ume {{\scriptstyle{\frac{1}{2}}}}
\def \ra {\rightarrow}

\def \eqq {\equiv}

\def \a {{\alpha}}
\def \b {{\beta}}

\def \d {{\delta}}

\def \om {{\omega}}

\def \A {{\cal A}}

\def \F {{\cal F}}
\def \G {{\cal G}}
\def \H {\mbox{${\cal H}$}}

\def \O {{\cal O}}
\def \P {{\cal P}}

\def \T {{\cal T}}

\def \Z {{\cal Z}}

\def \Psio {{\Psi_0}}

\def \d^nu {{\partial^\nu}}
\def \d^la {{\partial^\lambda}}
\def \d^o {{\partial^0}}

\def \x {{\bf x}}

\def \Nbf {{\bf N}}
\def \Rbf {{\bf R}}

\def\doppio#1{{\rm I}\kern-.1667em{\rm #1}}

\def\Q{\text{Q}\kern-.52em
    \text{\vrule height1.5ex width.5pt depth0pt}\kern.45em}

\def\dZ{{\mathchoice {\hbox{$\Ss\textstyle Z\kern-0.4em Z$}}
{\hbox{$\Ss\textstyle Z\kern-0.4em Z$}} {\hbox{$\Ss\scriptstyle
Z\kern-0.25em Z$}} {\hbox{$\Ss\scriptscriptstyle Z\kern-0.2em
Z$}}}}

\def\dC{{\mathchoice{\hbox{$\rm\textstyle\text{\kern.35em\vrule
   height1.5ex width.5pt depth0pt\kern-.35em C}$}}
{\hbox{$\rm\textstyle\text{\kern.35em\vrule
   height1.5ex width.5pt depth0pt\kern-.35em C}$}}
{\hbox{$\rm\scriptstyle\text{\kern.35em\vrule
   height1.5ex width.3pt depth0pt\kern-.35em C}$}}
{\hbox{$\rm\scriptscriptstyle\text{\kern.35em\vrule
   height1.5ex width.2pt depth0pt\kern-.35em C}$}}}}

\makeatletter

\@addtoreset{equation}{section}

\begin{document}

\maketitle
\begin{abstract}
The concepts of symmetry, symmetry breaking and gauge symmetries  are discussed,  their operational meaning being displayed by   the  observables {\em and}  the (physical) states.  For infinitely extended systems the  states fall into physically disjoint  {\em phases} characterized by their behavior at infinity or  boundary conditions, encoded in the ground state, which provide  the cause of   symmetry breaking   without contradicting  Curie Principle.  Global gauge symmetries,  not seen by the observables,  are nevertheless displayed by detectable properties of the states (superselected quantum numbers and  parastatistics).  Local gauge symmetries are not seen  also by  the physical states; they  appear only in non-positive representations of field algebras. Their role at the Lagrangian level  is merely to ensure the validity on the physical states of  local Gauss laws, obeyed by the currents which generate the corresponding global gauge symmetries; they are  responsible for most  distinctive physical properties of gauge quantum field theories. The topological invariants of a local gauge group define  superselected quantum numbers,  which account for the  $\theta$ vacua.

\end{abstract}

\newpage

\section{Introduction}

The concepts of symmetries, symmetry breaking and gauge symmetries, at the basis of recent developments in theoretical physics, have given rise to discussions from a philosophical point of view.\footnote{An updated and comprehensive account may be found in ~\cite{BC}.}
 Critical issues are the meaning of spontaneous symmetry breaking (appearing in conflict with the Principle of Sufficient Reason)  and the physical or operational meaning of gauge symmetries. 

The aim of this talk is to offer a revisitation of the problems strictly in terms of operational considerations. The starting point (not always emphasized in the literature) is the realization that 
  the description of a physical system involves both the {\em observables}, identified by the experimental apparatuses used for their measurements, {\em and}  the states, which define the experimental expectations.
Since the protocols of preparations of the  states may not always be  compatible, i.e. obtainable one from the other by physically realizable operations, the states fall into disjoint families, called phases, corresponding to incompatible realizations of the system. This is typically the case for infinitely extended systems, where different behaviors or boundary conditions of the states at space infinity identify disjoint phases due to the inevitable  {\em  localization} of any realizable operation. 

 This feature, which generically is not shared by finite dimensional systems, provides the explanation of the phenomenon of spontaneous symmetry breaking, since the boundary conditions at infinity  encoded in the ground state  represent the cause of the phenomenon in agreement with Curie principle.
   
The role of the states is also crucial for the physical meaning of gauge symmetries, which have been argued to be non-empirical because they are not seen by the observables. The fact that non-empirical constituents may characterize the theoretical description of subnuclear systems, as displayed by the extraordinary success  of the standard model of elementary particle physics, has provoked philosophical discussion on their relevance (see ~\cite{BC}). For the discussion of this issue it is important to distinguish global (GGS) and local gauge symmetries (LGS).\goodbreak

  The empirical consequences of the first is displayed by the properties of the states,  since invariant polynomials of the gauge generators define elements of the center of the algebra  of observables $\A$, whose joint spectrum labels the representations of $\A$ defining {\em superselected quantum numbers}; another empirical consequence of a global gauge group is the {\em parastatistics} obeyed by the states. Actually the existence of a gauge group can be inferred from such properties of the states. 

At the quantum level, the group of local gauge transformations connected to the identity  may be represented non-trivially only in unphysical non-positive  representations of the field algebra and therefore they reduce to the identity not only on the observables, but also on the physical states.  
   
From a  {\em technical} point of view, a role of LGS is to identify (through the pointwise invariance under them) the  {\em local} observable subalgebras of auxiliary field algebras (represented in non-positive representations). LGS also provide a  useful recipe for writing down Lagrangians which automatically lead to the  validity on the physical states  of {\em local Gauss laws} (LGL), satisfied by the currents which generate the corresponding GGS. Actually, LGL appear as the important  physical counterpart of LGS representing the  crucial distinctive features of Gauge QFT with respect to ordinary QFT.

A physical residue of LGS is also provided by their {\em local} topological invariants, which define elements of the center of the local algebras of observables, the spectrum of which label the inequivalent representations corresponding to the so-called $\theta$ vacua. The occurrence of such local topological invariants explains in particular the breaking of chiral symmetry in Quantum Chromodynamics (QCD), with no corresponding Goldstone  bosons.    

Finally, since   only observables {\em and} states ({\em identified} by their expectations of the observables \cite{FS1} \cite{FS2}) are needed for a {\em complete}  description of a physical system, and both have a deterministic evolution, the problem of  violation of determinism in gauge theories  looks rather an artificial issue  from a physical and philosophical  point of view. 

\newpage

\section{Symmetries and symmetry breaking}
For the clarification of the meaning and consequences of symmetries in physics,  from the point of view of general philosophy, a few basic concepts are 
helpful.

Quite generally, {\em the description of a physical  system} (not necessarily quantum!) is (operationally) given  \cite{FS1} \cite{FS2}
in terms of 

\noindent 1) the {\bf  observables}, i.e. the set of measurable quantities of the system, which characterize the system (and generate the so-called {\em algebra $\A$ of observables}) 

\noindent 2) their {\bf time evolution}   

\noindent 3) the set $\Sigma$ of physical {\bf states} $\omega$  of the system, operationally defined by  protocols of preparations and characterized  by their expectations of the observables $\{\omega(A), A \in \A\}$

Operationally, an observable $A$ is identified by the actual experimental apparatus which is used for its measurement, (two apparatuses being {\em identified} if they yield the same expectations on all the states of the system)

The first relevant point is  the {\em compatible realization of} two different {\em states}, meaning that they are  obtainable one from the other  by {\em physically realizable operations}. This  defines a partition of the  states into physically disjoint sets, briefly called {\bf phases}, with the physical meaning of describing disjoint realizations of the system, like disjoint thermodynamical phases, disjoint worlds or universes.  

For infinitely extended systems, in addition to the condition of {\em finite energy}, a very strong  physical constraint is that the physically realizable operations have inevitably some kind of {\em localization},   no action  at space infinity being physically possible.  Thus,  for the characterization of the states of a phase $\Gamma$,  a crucial role is played by their  large distance behavior or by the boundary conditions at space infinity, since they cannot be changed by physically realizable operations. Typically, such a behavior at infinity of the states of a given phase $\Gamma$ is codified by the lowest energy state or ground state  $\omega_0 \in\Gamma$, all other states of $\Gamma$ being describable as  ``localized'' modifications of it.  Thus, $\omega_0$ identifies $\Gamma$ and defines a corresponding (GNS) representation $\pi_\Gamma(\A)$ of the observables in a Hilbert space $\H_\Gamma$, with the cyclic ground state vector $\Psio$.\footnote {This point is  discussed for both classical and quantum systems in \cite{FS3}, \cite{FS4}.} \goodbreak

The simplest realization of  {\bf symmetries} is   {\em as transformations of the observables  commuting with  time evolution}, operationally  corresponding to the transformations of the  experimental apparatuses which identify the observables (e.g. translations, rotations). This is more general than Wigner definition of  {\em symmetries as transformations of the states  which leave the  transition probabilities invariant} (adapted to the case of the unique Schroedinger phase of atomic systems). 

Actually, the disentanglement of symmetry transformations of the observables (briefly {\bf algebraic symmetries}) from those of the states ({\bf Wigner symmetries}), is the crucial revolutionary step at the basis of the concept of spontaneous symmetry breaking, which comes into play when there is more than one phase. 

An algebraic symmetry $\b$  defines also a symmetry of the states of a phase $\Gamma$ (i.e. a  Wigner or {\bf unbroken symmetry}) iff it may be represented by unitary operators $U_\b$ in $\H_\Gamma$.
  
An algebraic  symmetry $\b$  always defines a symmetry  of the {\em whole} set of  states $\Sigma$: 
\be{\omega \ra \b^*\om \eqq \om_\b, \,\,\,\,\,\,\,\om_\b(A) \eqq \om(\b^{-1}(A)), \,\,\,\,\forall A \in \A,}\ee
but in general  $\om$ and $\om_\b$ need not  belong to the same phase $\Gamma$, i.e. their preparation may not be compatible, so that the   symmetry $\b$ cannot  be  experimentally displayed in $\Gamma$ as invariance of transition probabilities, by means of physically compatible operations ({\bf spontaneously broken symmetry}). Thus, the breaking of $\b$ in $\Gamma$ is characterized by  the existence of states $\omega \in \Gamma$ (typically  the ground or vacuum state $\omega_0$) such that $ \omega_\b \notin \Gamma$. 

The philosophical issue of symmetry breaking, also in connection with Curie principle, has been extensively debated often with misleading or wrong conclusions. 

A widespread opinion is that symmetry breaking occurs whenever the ground state is not symmetric, but this is not correct for finite systems,  for which (under  general conditions) there is only one (pure) phase $\Gamma$, so that both $\omega_0$ and $ \omega_{0 \,\b}$ belong to $\Gamma$ and  $\b$ is described by a unitary operator. 

Thus, the finite dimensional (mechanical) models,  widely used in the literature to illustrate  spontaneous symmetry breaking, on the basis of the existence of non-symmetric ground states, are conceptually misleading.\footnote{The standard models are a particle in a double well or in a mexican hat potential (see also \cite{G} \cite{L}). The example of an elastic bar on top of which a compression force is applied, directed along its axis, exhibits a continuous family of symmetry breaking ground states, but spontaneous symmetry breaking occurs only in the limit of infinite extension of the bar; otherwise, both in the classical as well in the quantum case, there is no obstruction for reaching one ground state from any other.}

On the other hand, for a {\em pure phase} of an  infinitely extended  system, thanks to the uniqueness of the translationally invariant state (implied by the cluster property which characterizes pure phases),  the non-invariance of the ground state $\omega_0 \in \Gamma $ {\em under an internal symmetry} $\b$ (i.e. commuting with space-time translations) implies 
that $\omega_{0\, \b}$ cannot belong to $\Gamma $ and  $\b$ is broken in $\Gamma$.  {\em Under these conditions},  the non-invariance of the ground state   provides an  explanation in agreement with Curie principle, identifying the cause in  non-symmetric boundary conditions at infinity encoded in the ground state (see \cite{FS3} pp.23, 102). The philosophically deep loss of symmetry requires the existence of disjoint realizations of the system, which is related to its infinite extension. 

The existence of an algebraic symmetry reflects on {\em empirical  properties of the states} and may  be inferred from them. In fact, an  unbroken symmetry implies the validity of Ward identities, which codify  the existence of conserved quantities and of selection rules satisfied by the states; for continuous symmetries the conservation laws hold even {\em locally} by the existence of current continuity equations implied by the first Noether theorem (\cite{FS4}, p.146-7).  
 For a continuous symmetry group $G$ broken in  $\Gamma$, even if the generators do not exist as operators in $\H_\Gamma$ ,  the existence of a representation of $G$ at the algebraic level, (\cite{FS3}, Chapter 15), implies {\bf symmetry breaking   Ward identities} which display corrections given by non-symmetric ground state expectations,  called non-symmetric order parameters; an important empirical consequence is the existence of Goldstone bosons, for sufficiently  "local" dynamics (\cite{FS3}, Chapters 15-17).      


\def \Rbf {{\bf R}}
\section{Global gauge symmetries}
For the debated issue of the empirical meaning of {\bf global gauge symmetries} (GGS) (which by definition act trivially on the observables),  a crucial (apparently overlooked) point is that a complete  complete description of a physical system involves {\em both} its algebra of  observables {\em and } the states or representations which describe  its possible phases. In fact, {\em even if} there is no  (non-trivial)  transformation  of the observables corresponding to GGS, GGS are  strictly related to  the existence of disjoint representations  of the observable algebra and their empirical meaning is to provide a  classification of them in terms of superselected quantum numbers \cite{DHR}.
This is clearly illustrated by the following examples.
 
\noindent {\bf Example 1}. Consider a  free massive fermion field $\psi$ transforming as the fundamental representation of an internal $U(2)= U(1) \otimes SU(2)$ symmetry  with the algebra of observables defined  by its pointwise invariance under $U(2)$. The existence of the (free) Hamiltonian selects the Fock representation  in  $\H_F$ for the field algebra $\F$ generated by $\psi$ and this implies the existence of  
the generator $N$ of  $U(1)$ and of  the Casimir invariant 
\be{ T^2 \eqq \sum_{\alpha =1}^3  (Q^\alpha)^2, \,\,\,\,Q^\alpha \eqq \int d^3 x \,\psi^*(\x) T^\alpha \psi(\x),}\ee 
with $T^\alpha$, $\a = 1,...3$, the representatives of the generators of $SU(2)$. $N$ and $T^2$  are  invariant under the  gauge group  $U(2)$ and  as such they (or better their exponentials $U_N(\a) = \exp{i \a N}, \,U_{T}(\b) = \exp{i \b T^2}$, $\a, \b \in \Rbf$) may be taken as elements of  the {\bf center $\Z$ of the observable algebra} $\A$.
The eigenvalues $n \in \Nbf$ of $N$ and $j (j+1)$ ($j \in \ume \Nbf$)  of $T^2$ label   the   representations of $\A$ in $\H_F$ and the  fermion fields $\psi^*, \,\psi$    act as  intertwiners  between the inequivalent representations  of $\A$, by increasing/decreasing the numbers $n$ and $j$.

Had we started by considering only the observable algebra $\A$, we would have found that its representations are labeled by the (superselected) quantum numbers  $n$ and $j(j+1)$, corresponding to the spectrum of the  central elements $U_N(\a), \,U_T(\b)$ and that  the state vectors of the  representations of $\A$ are obtained by applying intertwiners to the $n = 0, \,j =0$  representation, consisting of the Fock vacuum. \goodbreak

We would then be  led to consider a larger (gauge dependent) algebra $\F$ generated by the intertwiners, to  interpret $n$  as the spectrum of the generator $N$ of a $U(1)$ group and to infer the existence of an $SU(2)$ group with $j(j+1)$ the eigenvalues of the associated $T^2$. Such a reconstructed $U(2)$ group  acts non-trivially on the intertwiners, but trivially on the observables, namely is a global gauge group.

\noindent {\bf Example 2}. A familiar physical system  displaying the above structure is the quantum system of $N$ identical particles, even if in  textbook presentations the relation between  the gauge structure and the center of the observables is not emphasized. 

The standard treatment introduces the (Weyl algebra $\A_W$ generated by the) canonical variables of $N$ particles and,  by the very definition of  indistinguishability, the observable algebra $\A$ is characterized by  its pointwise  invariance under the {\em  non-abelian group $\P$ of permutations}, which is therefore a global gauge group.

As before, its role is that of providing a classification of the inequivalent representations of the observable algebra contained in the unique regular irreducible representation of $\A_W$, (equivalent to  standard Schroedinger representation)  in the Hilbert space $\H = L^2(d^{3 N} q)$, where $\P$ is unbroken. $\H $ 
decomposes into irreducible representation of the observable algebra, each being characterized by a Young  tableaux,  equivalently by the eigenvalues of the characters $\chi_i$, $i = 1, ...m$.\cite{D}
For our purposes, the relevant point is that the characters are invariant functions of the permutations and, as such,   may be considered as elements of the observable algebra, actually elements of its center $\Z$.  

Thus, as before, the gauge group $\P$ provides elements of the center of the observables  whose joint  spectra label the representations of $\A$  defining superselected quantum numbers. Beyond the familiar one-dimensional representations (corresponding to bosons 
 and fermions) there are  higher dimensional representations, describing {\bf parastatistics} (i.e. parabosons and parafermions).

Another empirical consequence of a global gauge group is  the ({\em observable}) statistics obeyed by the states, a parastatistics of order $d$  arising as the  result of an  unbroken (compact) global gauge group acting on ordinary (auxiliary) bosons/fermions fields \cite{DrHR}, \cite{H}. In the model of Example 1, 
an observable  consequence of the   global gauge group $U(2)$ is that the corresponding particle states are parafermions of order two (meaning that not more than two particles may be in a state). 
 The quarks have the properties of parafermions of order three as a consequence  of the color group $SU(3)$ 
 (historically this  was  one of its motivations).

In conclusion, contrary to the widespread opinion that the gauge symmetries are not empirical, the {\em global  gauge symmetries are displayed by the properties of the states} ({\bf superselected quantum numbers and parastatistics}) and actually can be inferred from them.\footnote{The empirical meaning of the invariant functions of the generators of a global gauge group has been pointed out in \cite{FS4}, pp.153-8 and later 
resumed by Kosso and others; (see also \cite{FS5}, Chapter 7).}

It must be stressed that  a global gauge symmetry  emerges as an empirical  property of a system by looking at the {\em whole set of its different realizations}; in a single factorial representation, the center of the observables is represented by a multiple of the identity and its physical meaning in terms of superselected
 quantum numbers is somewhat frozen. To reconstruct an  operator of the center of $\A$ one must look to its {\em complete spectrum}, i.e. to {\em all}  factorial representations of $\A$.

A continuous global gauge group  becomes particularly hidden in those representations in which the  exponentials of localized invariant polynomials of the generators  converge to zero when the radius of the localization region goes to infinity. This corresponds to the case in which, in the conventional jargon,  the {\bf global gauge group is broken}. 

In a representation $\H_\Gamma$ of the field algebra  in which the (continuous) gauge group $G$ is broken, briefly called a $G$-broken representation,   in contrast with  the above examples, the charged  fields do no longer intertwine between different representations of the observable algebra; in fact, they are obtainable  as weak limits of gauge invariant fields in the Hilbert space $\H_\Gamma$ ({\em charge bleaching}) \cite{MS}.

\noindent {\bf Example 4}. The   Bose-Einstein condensation  is characterized by  the breaking of a global  $U(1)$ gauge  group (acting on the Bose particle field as the $U(1)$ group of Example 1), as  very clearly displayed by the free Bose gas.\footnote{For a simple account see \cite{FS3},  p. 106.} 
The $U(1)$ breaking leads to the existence of {\bf Goldstone modes}, the so-called Landau phonons,  and the existence of such excitations may in turn indicate the presence of a broken $U(1)$ symmetry. 
     
Finally, the gauge group is also reflected in the {\em counting of the states}. In  $G$-unbroken representations of $\A$,  to each irreducible representation of $G$ contained in the field algebra $\F$,  there corresponds a  single physical state, whereas in the fully broken case to each $d$-dimensional irreducible representation   in $\F$,  there correspond $d$ different physical states \cite{FMS}  (for a handy account see  \cite{FS4}, Part B, Section 2.6).



\section{Local gauge symmetries}

Traditionally, a {\em local gauge symmetry} group is introduced as an extension of the corresponding global group $G$  by allowing the group parameters to become  $C^\infty$ functions of spacetime. It is however better to keep distinct the local gauge group $\G$ parametrized by strictly localized functions (technically of compact support) from the corresponding global one $G$, since the topology of the corresponding Lie algebras is very different and invariance under $\G$ does not imply invariance under $G$ (as displayed by the Dirac-Symanzik electron field, \cite{FS5}, p.159). 

Also from a physical point of view, the two groups are very different, since  in {\em any} (positive) realization  (of the system) the group of local gauge transformations connected with the identity is represented trivially, whereas the global gauge group displays its physical meaning through the properties of the states (see the above examples). For example,  the $U(1)$ global gauge group is non-trivially represented in Quantum Electrodynamics  (QED) by the existence of the charged states, whereas {\em the  local $U(1)$ group reduces to the identity on the physical states} (\cite{FS5},  Section 3.2). 

Therefore, the natural question is  which is the empirical meaning, if any, of a local gauge symmetry (LGS) $\G$ in QFT.  From a  technical point of view, pointwise invariance under $\G$ may be used for selecting the {\em local  subalgebra of observables}, from an auxiliary field algebra $\F$,  locality (strictly related to causality \cite{H})  not being implied by $G$ invariance (e.g. in QED $\bar{\psi}(x)\,\psi(y)$ is  invariant under $G = U(1)$,  but not under $\G$ and  is not a {\em local} observable field).  

A deeper insight on the physical counterpart of a LGS is provided by the second Noether theorem, according to which the invariance of the Lagrangian under  a group of local gauge transformations $\G$ implies that the currents which generate the corresponding global group $G$ are the divergences of  antisymmetric tensors  
\be{ J_\mu^a(x) = \partial^\nu \,G_{\nu \mu}^a(x)\,\,\,\,\,\,\,G^a_{\mu\, \nu} =  - \,G^a_{\nu \,\mu}.}\ee 
({\bf local Gauss law} ).

This is  a very strong constraint on the physical consequences of $G$ (corresponding to the Maxwell equations in the abelian case).  Actually, such a property seems to catch the essential consequence of local gauge symmetry, since $\G$ invariance of the Lagrangian is destroyed by the gauge fixing, whereas  the corresponding local Gauss laws (LGL) keep  holding on the physical states, independently of the gauge fixing.\footnote{A gauge fixing  which breaks the global group $G$  involves a symmetry breaking order parameter and it is consistent only if $G$ is broken (see \cite{FS5}, p. 178 and \cite{DPS}).}

Moreover,  a LGL implies that $\G$ invariant  {\em local} operators  are also  $G$ invariant. In the abelian case this implies the {\bf superselection  of the electric charge} (\cite{FS5}, Sect.5.3)

Thus, it is  tempting to downgrade  local gauge symmetry to a   merely  technical recipe for writing down Lagrangian functions, which automatically lead to  LGL for the currents which generate the corresponding  global gauge transformations. \footnote{ The fact that LGL represent the distinctive physical property of "local gauge theories" has been discussed and emphasized in \cite{FSG}, \cite{FS4}, p.\,146-149, and later rediscovered, without quoting the above references.}

The physical relevance of a LGL is that it encodes a general property largely independent of the specific Lagrangian model and in fact, most of the peculiar (welcome) features of Gauge QFT, with respect to standard QFT,  may be shown to be direct consequences of the validity of LGL (see \cite{FS5}, Chapter 7):

\noindent a) a LGL law implies that {\em states carrying} a (corresponding) {\em global gauge charge cannot be localized}; this means that the presence of a  charge  in the space time region $\O$ can be detected  by measuring observables localized in the (spacelike) causal complement $\O'$; this represents a very strong departure from standard QFT, where ``charges'' in $O$ are not seen by the observables localized in $\O'$;

\noindent b) LGL provide direct explanations of the evasion of the Goldstone theorem   by global gauge symmetry breaking (Higgs mechanism);

\noindent c) particles carrying a gauge charge (like the electron) cannot have a sharp mass ({\em infraparticle phenomenon}), so that they are {\em not Wigner particles};

\noindent d) the non-locality of the ``charged'' fields, required by the Gauss law,  opens the possibility of their failure of satisfying the cluster property with the possibility of a linearly raising potential, as displayed by the quark-antiquark interaction, otherwise precluded in standard QFT (where the cluster property follows from locality); 

\noindent e) a local gauge group may have a non-trivial topology, displayed by components disconnected from the identity, and the corresponding {\em topological invariants}
defines elements of the center $\Z$ of the local algebra of observables $\A$; for Yang-Mills theories such elements $\T_n(\O)$, localized in $\O$, are labeled by the winding number $n$ and define an abelian group ($\T_n(O) \T_m(O) = \T_{n+m}(O)$);  their spectrum $\{ e^{i 2 \pi n \theta}, \,\,\theta \in [ 0, \pi)\,] \}$ labels the factorial representations of the local algebra of observables, the corresponding ground states being the {\em  $\theta$-vacua}. They are unstable under the chiral transformations of the axial $U(1)_A$ and therefore chiral transformations are inevitably broken  in {\em any } factorial representation of $\A$ without Goldstone bosons. Thus, the topology of $\G$ provides an explanation of chiral symmetry breaking in QCD, without recourse to the instanton semiclassical approximation (\cite{FS5}, Chap.\,8).  

In conclusion, LGS are not symmetries of nature in the sense that they reduce to the identity not only on the observables, but also on the states, possibly except for their local topological invariants. From the point of view of general philosophy, they appear in Gauge QFT as merely technical devices to ensure the validity of local Gauss laws (through a mathematical path which uses an invariant Lagrangian {\em plus} a non-invariant gauge fixing). 

By the same reasons, i.e. the realization that the observables and the physical states are the only  quantities needed for the complete description of a physical system, the issue of  violation of determinism in gauge theories  does  not deserve physical and philosophical attention, since the observables and the physical states have a deterministic time evolution.  




\section{Additional discussion required by  the referee}
The aim of the paper is to present logical (mathematically sound) arguments and critical discussion of ideas and proposals which were previously not sufficiently elaborated from a philosophical point of view; 
in particular the paper  aim is to criticize misleading or wrong conclusions drawn from eminent philosophers of physics.

\vspace{2mm}
\noindent {\em Empirical meaning of symmetries} 

For the discussion of the empirical meaning of symmetries it is important to take into account the basic result of (the first) Noether theorem, by which invariance (of the dynamics) under a continuous one-parameter group of transformations is equivalent to  the existence  of a conserved quantity; hence, the empirical meaning of a symmetry may be provided by the empirical realizations of the  symmetry transformations (e.g. space translations, rotations etc.)  {\em as well as} by the empirical meaning of the associated conserved quantity, which represents the generator of the symmetry. Thus, e.g. the empirical meaning of space translations may be argued by the actual operational realizability of such transformations (in terms of translating observable quantities), as well as by the empirical meaning of the (observable) conserved space momentum. Therefore,  it is not appropriate  to regard the second manifestation as of {\em indirect} empirical significance (as stated in \cite{BB}), since from an experimental point of view this is by far the more easy way for detecting the existence of a symmetry, as also argued by Morrison \cite{M} : "Conservation laws provide the empirical component or manifestation of symmetries". 

The peculiarity of a global gauge symmetry is that it cannot be realized as a group of transformations of the observables (being  the identity on them), but nevertheless the associated conserved quantity may have an empirical significance in terms of empirical properties of the states, as it is clearly displayed in Quantum Electrodynamics (QED), where the generator of global gauge transformations describes the electric charge of the states, a very relevant conserved physical property. We therefore essentially adopt the following criterium for empirical significance, stated by Earman \cite{E}: "What is objective or real in the world is described by the behavior of the values of genuine physical magnitudes of the theory", however with the crucial gloss   that genuine physical quantities include {\em both} the observables {\em  and} the states of the given physical systems. 

In conclusion, a symmetry has an empirical significance if it is displayed by properties of the observables (e.g. by defining automorphisms of the algebra of observables) or of the physical states (e.g. by providing conserved quantum numbers which classify the states). It follows that global gauge symmetries are empirical, since their generators provide the conserved superselected quantum numbers which label the physical states, but generally  local gauge symmetries  are not.
To my knowledge, the above relevant gloss has been missed in the discussions on the empirical significance of gauge symmetries, even in papers aiming to clarify the philosophical aspects of gauge symmetries \cite{F}.
   
\vspace{2mm}
\noindent {\em Empirical meaning of local  gauge symmetries} 

Practically the whole morning section of the meeting (during which the present paper was presented) was occupied by talks centered on the possible  philosophical meaning of local gauge symmetries, dwelling on the philosophical meaning of invariance under local transformations which reduce to the identity on the observables. As argued in Section 4, this looks like a metaphysical issue and, as such,  does not deserve scientific attention. The distinction between global and local gauge symmetries is crucial for the discussion of the empirical meaning of gauge symmetries  since only the first have a physical meaning whereas  local gauge transformations do not. 

To this purpose, I quote the final conclusion by Elena Castellani in her contribution "Symmetry and equivalence" in "Symmetries in Physics" (Ref.1): "Today we believe that global gauge symmetries are unnatural...We now suspect that all fundamental symmetries are local gauge symmetries". In the same book, in the conclusion of his contribution "The interpretation of gauge symmetry" M. Redhead writes "The gauge principle is generally regarded as the most fundamental  cornerstone of modern theoretical physics. In my view its elucidation is the most pressing problem in current philosophy of physics".
  
For the discussion of this problem it is crucial to keep distinct the group of gauge  transformations which 
differ from the identity only on compact bounded regions, henceforth called {\em local},  and the gauge group of {\em global} (i.e. independent from the point in space time) transformations; englobing both under the name of a local gauge group is, in my opinion, not convenient and likely misleading, because it hides the fact that they have a different status about empirical significance and, moreover, invariance under localized gauge transformations does not imply invariance under the corresponding global ones. Hence, as argued in my paper, the two groups should be taken neatly in separate boxes.  

Then, the interesting question is what is the role of local gauge symmetries (equivalently of the gauge principle) in the constructions of models of elementary particles and the answer discussed in Section 4 is that they enter only as intermediate steps, doomed to lose any operational and philosophical meaning at the end (except for the related topological invariants, see below). Their merely intermediate role is to lead to the formulation of a dynamics characterized by the validity (on the physical states) of {\em local Gauss laws} obeyed by the currents which generate the corresponding global gauge symmetries. Such Gauss laws are not spoiled by the inevitable gauge fixing, needed for quantization (the proof of their validity on the physical states is not trivial in general \cite{DPS}, even if it is out of discussion in QED): they are detectable  properties of the physical  states and, as discussed in Section 4, they provide the physical and philosophical distinctive characterization of gauge quantum field theories. 
  
This pattern is clearly displayed by Quantum Electrodynamics  where (one may prove that): 1) the local gauge group reduces to the identity both on the observables as well on the physical states, i.e. does not have any empirical meaning, 2) on the other hand, the local Gauss law (somewhat related to the {\em intermediate} use of the non-empirical local gauge invariance)  has an empirical significance, being one of the Maxwell equations, 3) the global gauge group has an empirical meaning, since its generator is the electric charge, whose corresponding quantum number is superselected.
  
The recognition that local Gauss laws  are the characteristic features of gauge quantum field theories has been argued and stressed in view of quantum theories in \cite{SW} \cite{FSG} \cite{FS4} and later reproposed, without quoting the above references, by Karatas and Kowalski (1990) \cite{KK}, Al-Kuwari and Taha (1990) \cite{AT}, Brading and Brown (2000) \cite{BBG}. Actually, such papers confine the discussion to the derivation of local Gauss laws from local gauge invariance (second Noether theorem at the {\em classical level, with no gauge fixing}), missing the crucial fact that at the quantum level local gauge invariance of the Lagrangian has to be broken by the gauge fixing and it is devoid of any empirical (and philosophical) significance, whereas the validity of local Gauss laws keeps being satisfied by the physical states, and it explains the interesting (revolutionary) properties of gauge theories  (as explained in Section 4). 

 In contrast with global gauge symmetries, local gauge symmetries are only useful tricks used in {\em intermediate} steps (which use an auxiliary unphysical field algebra, initially a Lagrangian which has local gauge invariance, to be next broken by the gauge fixing, a redundant space of vector "states", only a subspace of which describes physical states, on which local gauge symmetries reduce to the identity). The final emerging picture is a description of the physical system characterized by conserved (actually superselected) quantum numbers,  provided by the generators of the global gauge symmetry, and by the validity of local Gauss laws (no trace remaining of local gauge invariance).

In my opinion,  from a philosophical point of view, one  should invest in the meaning of local Gauss laws rather than on local gauge invariance (or on the so-called Gauge Principle).

\vspace{2mm}  
\noindent{\em Determinism}
  
The issue of violation of determinism should not even be raised, being discussed with reference to equation of motions for gauge dependent variables which  are deprived of objectivity and of reality, the objective description of a physical system involving only (the properties of)  observables {\em and}  physical states, whose time evolution is deterministic. 
   
   Quite generally, all what is needed for the complete description of a physical system is the determination of the time evolution of its observables and states, but for the solution of the related mathematical problem 
one may use tricks and  auxiliary variables in intermediate steps for which there is no need of a physical (and philosophical) interpretation. Only the final goal and result is relevant and there is a plenty of examples of such a technical strategy in theoretical physics.  Thus, in gauge theories it is technically convenient  to introduce an auxiliary (gauge dependent) field algebra with well defined dynamics, i.e. such that the (mathematical) Cauchy problem  for its time evolution is well posed (existence and uniqueness of solutions).  To this purpose one has to introduce  a gauge fixing in the Lagrangian, even if it is not necessary to completely fix the gauge; e.g. the Cauchy problem has been proved to be well posed in the Feynman-Gupta-Bleuler gauge, in the temporal gauge, in the Lorentz gauge (all allowing a residual symmetry group of non-constant gauge transformations).  The observables are characterized as the functions  of such auxiliary fields which are invariant under local gauge symmetry and satisfy locality; this is the (merely) technical role of local gauge symmetry. 

In quantum mechanics, once the Hamiltonian $H$ has been defined (as a self-adjoint operator) the time evolution is described by the unitary one-parameter group generated by $H$ and therefore the time evolution is automatically deterministic; thus, for field  quantization only those field operator may be introduced which have a deterministic evolution. This is why the quantization of gauge theories requires the introduction  of a gauge fixing such that the initial value problem of the (auxiliary) field algebra has a unique solution. 

\vspace{2mm}
\noindent {\em Infinitely extended systems and SSB}

In order to be (spontaneously) broken,   a symmetry,   defined as an automorphism/transformation of the observables, must fail to be implementable by unitary operators acting on the states of a physical realization of the system (otherwise one has an unbroken, i.e. Wigner symmetry). This is possible only if there exist   
disjoint realizations of the system (with the meaning of disjoint phases or worlds) all described by the same algebra of observables with the same  time evolution. The physical/empirical meaning of disjointness is that configurations or states of the system belonging to different phases cannot be prepared in the same laboratory, more generally their protocols of preparation are not compatible. In mathematical language this amounts to the impossibility of describing states of different phases by vectors of the same Hilbert space carrying an irreducible or factorial representation of the algebra of observables. SSB in one realization or phase is explained by, and actually  equivalent to, the instability of the phase under the symmetry, by the reason that in order to empirically detect the existence of a symmetry one must be able to operationally compare the behavior of each given configuration with that of its transformed one.

 For quantum systems described by a finite number of canonical variables  (under general regularity conditions, by Stone-von Neumann theorem) there is one phase and therefore no SSB, even if there are non-symmetric ground states,
in contrast with the wrong conclusion drawn from classical finite dimensional models with  non-symmetric ground states.
This leaves open a possibility for  systems described by an infinite number of canonical variables, in particular for infinitely extended systems (which require an infinite number of canonical variables). 

Then, the next issue is the existence of disjoint phases for infinitely extended systems; in this case different behaviors or different boundary conditions at space infinity of  configurations (or states) of the system imply that their preparations are not compatible, since the inevitable localization of any physically realizable operation (involved in passing from one preparation to another) precludes to change the behavior at infinity. Hence, generically infinitely extended systems exhibit more than one phase, characterized by the boundary conditions at infinity, which are generally  encoded in the ground state of the given phase, see Proposition 6.3 of Ref.4)  and SSB  may occur. 

In conclusion, the crucial ingredient for symmetry breaking is the existence of disjoint phases and this occurs for  infinitely extended systems (though not exclusively).   
  
\vspace{2mm}
\noindent {\em References}

One of the referee request was to comment on a list of papers dealing with overlapping subjects, qualifying the  novelties  (if any)  with respect to them, (a task, which I  will reluctantly try).  
    
\noindent 1) {\em Brading and Brown \cite{BB}}. As in all papers by philosophers of physics, which I know of, the discussion overlooks the important fact that an objective description of a physical system should exclusively be based on (the properties of the) observables {\em and} states and that the empirical significance of symmetries should be argued in such terms (e.g. automorphisms of the observables and/or conservation laws obeyed by the states, as explained above). The missing clear distinction of global versus local gauge symmetries  precludes  to immediately reach the conclusion about the empirical significance of the former
and the {\em impossible} empirical significance of the latter. In fact, in that paper  local symmetries are identified as those which depend on "arbitrary smooth functions of space and time"; the lack of any  localization restriction implies that the so defined group of local symmetries contains  the  group of global symmetries as a subgroup, since, as every first year student in  mathematics knows, the constant functions satisfy the  smoothness  condition (the excuse that localizability was tacitly
assumed  would mean a lack of precision without which
 mathematics as well as logic do no longer exist).
 
 Had Brading and Brown   clearly understood the different status
of the two groups and the general argument that local gauge symmetries reduce to the identity both on the
observables as well as on the states, they might have reduced their
paper to a few lines.

\vspace{2mm} \noindent 2) {\em Healy 2010 \cite{He}}. 
The paper  looks as  a rather sketchy account of the common (heuristic) wisdom  about $\theta$ vacua, completely ignoring the critical revisitation of such a subject, presented  in \cite{MSQCD} and later further discussed in Ref. [13]. In my opinion, this is not merely a question of mathematical physics precision, since it is very dangerous and certainly not satisfactory to ground a philosophical discussion on ideas, which may have a useful heuristic value, but have serious problems of mathematical and {\em logical} consistency. This applies to Healy paper, as I shall try to explain. 

The winding number $n$ defined in eq. (10), a crucial ingredient of the discussion, requires that $A_i(x)$ are continuous functions and therefore it looses any meaning for relativistic quantum fields, which have been proved to be singular "functions" of space points (technically operator valued tempered distributions). In fact, in order to give a possible meaning to such an equation the standard theoretical physics wisdom is to apply it to  regular  (euclidean) field configurations in the functional integral formulation (of quantum field theory), the so-called instantons. However, continuity is required and continuous euclidean configurations have zero functional measure (this problem is well known to the eminent theoretical physicists  who contributed to this subject, like Coleman, Weinberg etc.). This consistency problem was solved in \cite{MSQCD} in a way that has strong philosophical consequences; in fact, no reference is made to the  topological structure of the  (questionable) semiclassical instanton approximation (of the functional integral) and the proposed solution exclusively exploits the  topological invariants of the (non-abelian) local gauge group.  It is shown that such topological  invariants define elements of the center of the local observable algebra and their spectrum (i.e. the $\theta$ angle) characterize the $\theta$ vacua. From a general philosophical point of view, the conclusion is that even if the (group of) local gauge transformations connected with the identity reduce to the identity both on the observables as well as on the physical states, the topological invariants which classify the other components disconnected from the identity provide detectable superselected quantum numbers 
(the $\theta$ angles), which classify the physical states, just as the generators of  global  gauge group do. In conclusion, {\em local gauge symmetries are not empirical except for their topology}.

 The first sentence of the paper, with the abstract definition of a symmetry as "an automorphism-transformation that maps the elements of an object onto themselves so as to preserve the structure  of that object" is too loose and imprecise. Which elements (observables? states?)? For the states a symmetry may possibly  preserve the relations between them (preserving transition probabilities); "to preserve the structure" does not have a sharp clear meaning. This applies also to the subsequent attempt of formalization (A 1-1 mapping $\phi : S \rightarrow S$ of a set of situations...) which uses an undefined (vague) concept ("situations"). 

The merely intermediate role of local gauge symmetries for the validity of local Gauss laws has been missed.

At the end of Section 3. The last two statements are rather misleading. First, local gauge transformations, as well as the topological invariants provided by  them, do not relate configurations associated to different vacua; rather the topological invariants define elements of the center of the observables which label (not relate!) the vacua. The author seems to overlook the crucial difference between the empirical significance of a symmetry displayed by transformations or relations (between observables or states) and the empirical significance displayed by the existence of conservation laws (as argued by Morrison).  Similarly, the statement at the end of Section 4, that "a large gauge transformation represents a change from one physical situation to another" is conceptually wrong.

Towards the end of Section 5. The "generator" $\hat{U}$ of a large gauge transformation cannot be defined because the group of large gauge transformation is not continuously connected with the identity; it is a mathematical non-sense. What may be defined, as done in \cite{MSQCD}, are the elements $T_n$ of the quotient $G/G_0$ of the local gauge group $G$ with the local group $G_0$ of transformations connected with the identity (having zero winding number). Such a quotient is an abelian group, whose elements belong to the center of the local observable algebra  and their spectrum (or eigenvalues) are the $\theta$ angles.

The paradox raised at the beginning of Section 6: "a global gauge transformation appears as a special case of a large gauge transformation" is a consequence of the improper choice of not distinguishing global and local gauge transformations (see above discussion).

\vspace{2mm} \noindent 3) {\em Struyve 2011 \cite{S}}. The paper is confined to discussing classical field theories, which are known to have serious problems about their physical interpretation, in particular for  elementary particles  interactions; they may provide some heuristic mathematical information, but they {\em  do not describe nature}, (with the possible exception of classical gravity, which however requires quantum effect for the description of black holes). The most objectionable  point is the  discussion of  SSB in terms of small perturbations around a non-symmetric ground state. As discussed in Ref. 4, in classical field theory, the set of small perturbations around the ground state solution is not stable under time evolution and therefore it looses  meaning with the passing of time. The set of "perturbations" of a ground state solution $\phi_0$, which are stable under time evolution are   
 those which define  a Hilbert sector or a phase, and are  of the form $\phi_0 + \chi$, with $\chi \in H^1$, $\partial_t \chi \in L^2$ (the corresponding theorems are discussed in \cite{FS3};  neither $\chi$ nor $\dot{\chi}$ remain small! SSB cannot be identified with the instability under the symmetry  of the set of small perturbations ("When considering small perturbations around a particular ground state, the equations of motions will not posses the symmetry of the fundamental equations of motion and one speaks of SSB.", at the beginning of Section 2.2.). 
The widespread cheap heuristic account/explanation of SSB in terms of small perturbations around  a non-symmetric ground state is not (mathematically) correct (as discussed in \cite{FS3}).

Last but not least, I do not see what the paper significantly add to the gauge invariant account  for the Higgs mechanism, in the full quantum case, given  by Frohlich-Morchio-Strocchi \cite{FMS}, which does not even appears in the references of Struyve paper.

\vspace{2mm} 
\noindent 4) {\em Smeenk 2006, \cite{SM}}. The paper is well written, but most of the general discussion of conceptual problems  is not novel and largely taken from \cite{FS3} \cite{FS4}. 
  
  The aim of the paper, stated in the Abstract and in the Introduction ("This article focuses on two problems related to the Higgs mechanism... what is the gauge invariant content of the Higgs phenomenon? and what does it means to break a local gauge symmetry?") is superseded  by \cite{FMS}, quoted only at the very end, probably to comply a referee request. The logical and conceptual discussion of the problems of the Higgs mechanism, together with their solutions, already appeared  in \cite{FS4} and in the 2005 edition of \cite{FS3}, which are not even mentioned in the references. E.g. the discussion of SSB in Section 2 heavily relies on  \cite{FS3}, in particular for SSB  in classical theories, for the exclusion of SSB in finite-dimensional quantum systems by Stone-von Neumann theorem, for the role of  the infinite extension  for SSB in spin systems. The content of footnote 5 is somewhat misleading, since {\em both} in Statistical Mechanics (SM) {\em as well as} in Quantum field theory in order to witness  SSB one must consider pure phases, i.e. ground state representations which satisfy the cluster property (this may require a decomposition of the representation obtained in terms of the partition function   in SM or of the functional integral in QFT). 

    In Section 3, the discussion of the Goldstone theorem and the crucial role of locality, usually overlooked in textbook treatments, relies on \cite{FS3}, Chapter 15, especially Section 15.2. The general non-perturbative proof that in local gauges the Goldstone bosons cannot be physical was given in \cite{FSH}, \cite{FS3}, Theorem 19.1, again not even quoted; the evasion of the Goldstone theorem in the Coulomb gauge due to   the lack of locality (rather than the lack covariance) is again clearly discussed in the 2005 edition of \cite{FS3}. 
The discussion of Elitzur theorem and its consistency with the occurrence of symmetry breaking  in several gauges (like e.g. the Coulomb gauge) was clarified in \cite{MS} and discussed at length in \cite{FS4}, Part C, Chapter II, 2.5, so that the discussion in Section 5 of Smeenk paper does not seem to add anything new.

\end{document}